\documentclass[reprint,preprintnumbers,amsmath,amssymb,nofootinbib,aps,prx]{revtex4-1}

\usepackage{slashed}
\usepackage{amsmath}
\usepackage{amssymb}
\usepackage{amsthm}
\usepackage{hyperref}
\usepackage{indentfirst}
\usepackage{psfrag}
\usepackage{graphicx}
\usepackage{cleveref}
\usepackage[utf8]{inputenc}

\hypersetup{colorlinks=true, linkcolor=blue, urlcolor=blue, citecolor=blue, linktocpage=true}

\usepackage{xifthen}
\newcommand*\diff{\mathrm{d}} 
\newcommand*\ldiff[2][]{ \ifthenelse{\isempty{#1}}{ \diff #2}{\diff^#1#2} \,} 
\let\limitint\int 
\renewcommand{\int}{\limitint \!} 

\def\e{{\rm e}}
\def\d{\partial}

\newcommand{\bseq}{\begin{subequations}}
\newcommand{\eseq}{\end{subequations}}

\newcommand{\arctg}{\mathop{\rm arctg}\nolimits}

\renewcommand{\th}{\mathop{\rm th}\nolimits}

\newcommand{\sh}{\mathop{\rm sh}\nolimits}
\renewcommand{\ln}{\mathop{\rm ln}\nolimits}

\newcommand{\bra}[1]{\langle #1 |}
\newcommand{\ket}[1]{| #1 \rangle}

\newcommand{\be}{\begin{equation}}
\newcommand{\ee}{\end{equation}}
\newcommand{\beqa}{\begin{eqnarray}}
\newcommand{\eeqa}{\end{eqnarray}}
\renewcommand\l{\lambda}
\newcommand\vf{\varphi}

\newcommand{\arcsh}{\mathop{\rm arcsh}\nolimits}

\renewcommand{\O}{\Omega}
\newcommand{\Tau}{\mathcal{T}}
\newcommand{\s}{\sigma}

\begin{document}

\vspace{-4.0cm}
\begin{flushright}
{\small FTPI-MINN-22-28,  UMN-TH-4203/22} 
\end{flushright}
\vspace{0.5cm}

\title{On thermal false vacuum decay around black holes }

\author{Vadim Briaud$^1$}
\email{vadim.briaud@phys.ens.fr }

\author{Andrey Shkerin$^2$}
\email{ashkerin@umn.edu}

\author{Sergey Sibiryakov$^{3,4}$}
\email{ssibiryakov@perimeterinstitute.ca}

\affiliation{$^1$Laboratoire de Physique de l’\'Ecole normale sup\'erieure,  ENS, Universit\'e PSL, CNRS,\\Sorbonne Universit\'e, Universit\'e Paris Cit\'e, F-75005 Paris, France,
$^2$William I. Fine Theoretical Physics Institute, School of Physics and Astronomy, University of Minnesota, Minneapolis, Minnesota 55455, USA\\$^3$Department of Physics \& Astronomy, McMaster University, Hamilton, Ontario, L8S 4M1, Canada\\$^4$Perimeter Institute for Theoretical Physics, Waterloo, Ontario, N2L 2Y5, Canada }

\begin{abstract}

In flat space and at finite temperature, there are two regimes of false vacuum decay in quantum field theory.
At low temperature, the decay proceeds through
thermally-assisted tunneling described by periodic Euclidean
solutions --- bounces --- with non-trivial time dependence. On the other hand, at
high temperature the bounces are time-independent and describe 
thermal jumps of the field over the potential barrier. We argue that
only solutions of the second type are relevant for false vacuum decay
catalyzed by a black hole
in equilibrium with thermal bath.
The argument applies to a wide class of spherical black holes,
including $d$-dimensional AdS/dS-Schwarzschild black holes and
Reissner--Nordstr\"om black holes sufficiently far from criticality. 
It does not
rely on the thin-wall approximation and 
applies to
multi-field scalar theories.
\end{abstract}

\maketitle

\section{Introduction}

False vacuum decay in field theory has been subject of extensive research, see \cite{Weinberg:2012pjx} for a
review.
In flat spacetime at zero temperature, the transition from the false to true vacuum
proceeds
via tunneling through the potential barrier separating the vacua.
In the semiclassical approximation, the tunneling is described by
bounce---regular solution of the field equations that
satisfies vacuum boundary conditions. 
The latter are conveniently formulated in the purely imaginary
(Euclidean) time~$\tau$~\cite{Coleman:1977py,Callan:1977pt}. 

At finite temperature $T$, the tunneling is described by thermal
bounce which is periodic in $\tau$ with the period $1/T$
\cite{Linde:1980tt,Linde:1981zj}. 
The thermal bounces depend on $\tau$ below certain critical
temperature, $T<T_c$, in which case they are also called `periodic
instantons'. At higher temperatures the bounces degenerate into $\tau$-independent
solution. 
Every constant-$\tau$ slice of this solution coincides with sphaleron (or critical bubble), which is the saddle-point configuration at the top of the
barrier separating the false and true vacua. 
This picture admits a 
natural physical interpretation. While at $T<T_c$ the vacuum decay
proceeds via quantum tunneling from an excited state, at $T>T_c$
thermal fluctuations of the field are strong enough to drive the
system over the barrier classically.  
Unless $T$ is very large, the decay rate is exponentially suppressed, 
\be
\Gamma_{\rm decay}\sim \e^{-B_\lambda}\;,
\ee
and the suppression factor $B_\l$ equals the Euclidean action of the
theory evaluated on the bounce solution. At $T>T_c$ this is simply
given by the sphaleron energy divided by $T$.

\begin{figure}[t]
	\center{	
		\begin{minipage}[h]{0.99\linewidth}
			\center{\includegraphics[width=0.9\linewidth]{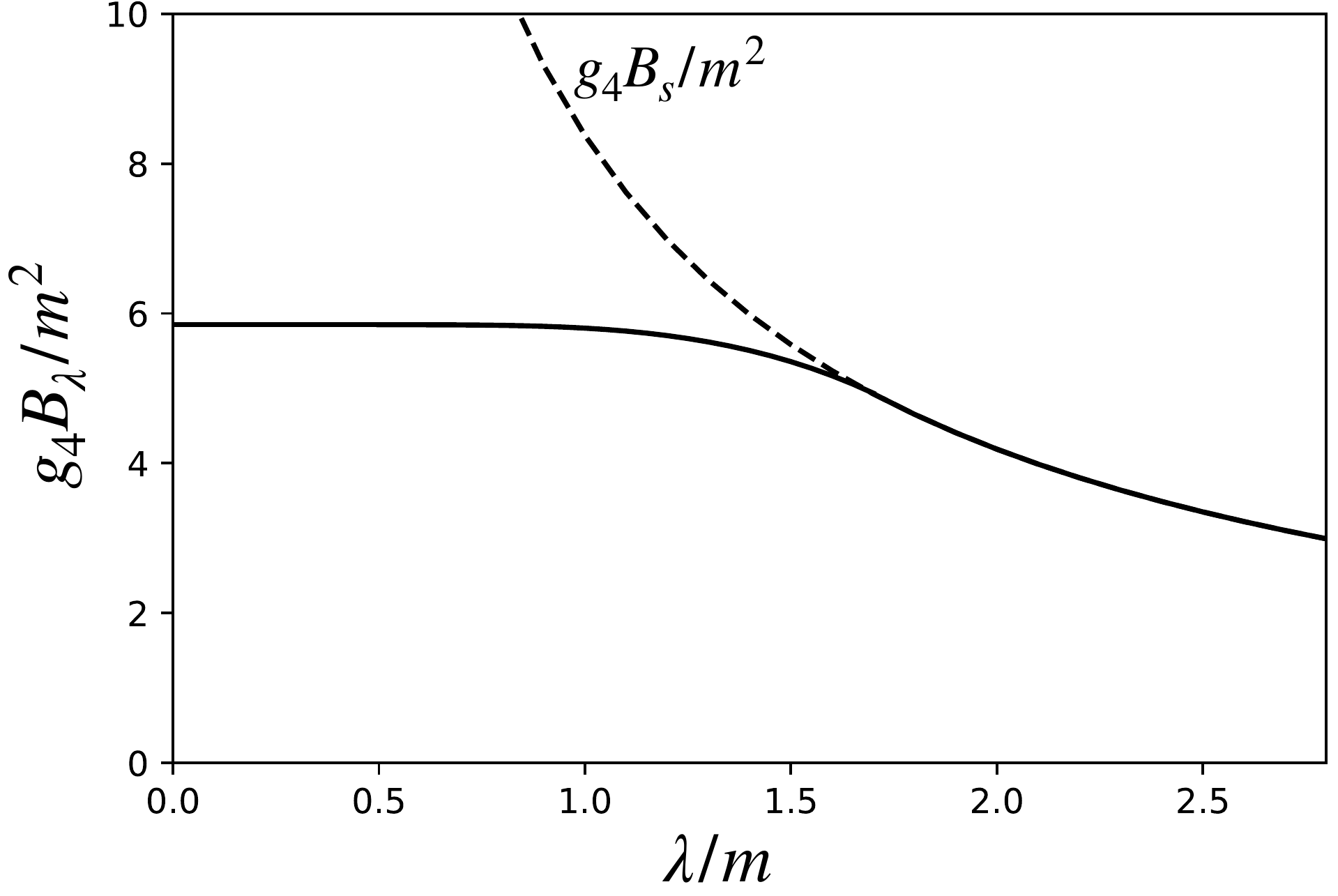}}
		\end{minipage}
	}
	\caption{Exponential suppression of false vacuum decay in
          $2$-dimensional 
flat spacetime at finite temperature $T\equiv \l/(2\pi)$.
	The dashed line shows the suppression $B_s$ due to sphalerons.
	We consider the scalar field theory with the potential
        (\ref{V_2d}).}
	\label{fig:2d_B_flat}
\end{figure}

We illustrate this on fig.~\ref{fig:2d_B_flat} where we plot $B_\l$ in
the $2$-dimensional theory of the real scalar field $\vf$ with the potential
\begin{equation}\label{V_2d}
V=\frac{m^2\vf^2 }{2}-\frac{\rm{g}_4\vf^4}{4} \;,
\end{equation}
where $\rm{g}_4>0$.\footnote{The model does not have a stable true
  vacuum since the potential is unbounded from below at
  $\vf\to\infty$. This does not affect our discussion.}
In this theory the periodic instantons smoothly merge to the sphaleron
branch at $T_c\sim m$. Though the $\tau$-independent bounces are still
solutions of the field equation at $T<T_c$, their Euclidean action is larger
than that of the periodic instantons. One can understand that they are
not relevant for vacuum decay by counting the number of their negative
modes --- linearly independent perturbations reducing their Euclidean
action. A relevant bounce, being a saddle point of the
action, must have exactly one such mode
\cite{Coleman:1977py,Callan:1977pt}. On the other hand, the
$\tau$-independent bounces at $T<T_c$ have at least two. One of them
corresponds to the change of the sphaleron size and exists also at
$T>T_c$. Another one appears only at $T<T_c$ and corresponds
to the deformation into periodic instanton.   

In general, the transition between periodic instantons and
$\tau$-independent bounces can be discontinuous, as e.g. in models
admitting the thin-wall approximation in spacetime with $d>2$
dimensions \cite{Garriga:1994ut}. Still, the presence of extra
negative modes around $\tau$-independent bounces at low temperature
persists and provides a clear signal that these solutions are
irrelevant for flat-space vacuum decay \cite{Weinberg:2012pjx}.

The goal of the present paper is to understand how the above picture of
thermal vacuum decay is affected by the curved metric of a black hole
(BH). It is known that a BH catalyzes the decay providing a nucleation
site for bubbles of the true vacuum 
\cite{Hiscock:1987hn,Berezin:1987ea,Arnold:1989cq,Berezin:1990qs,
Gregory:2013hja}. Various aspects of this effect and its possible
relevance for the instability of the electroweak vacuum of the
Standard Model have been discussed in 
\cite{Sasaki:2014spa,Burda:2015isa,Burda:2015yfa,Burda:2016mou,
Tetradis:2016vqb,Chen:2017suz,Gorbunov:2017fhq,Canko:2017ebb,Mukaida:2017bgd,
Kohri:2017ybt,Hayashi:2020ocn,Miyachi:2021bwd,Shkerin:2021zbf,Shkerin:2021rhy,
Strumia:2022jil}. Assume that a BH with Bekenstein--Hawking temperature $T_{\rm
BH}$ is surrounded by thermal bath with the same temperature. This
equilibrium state is known as the 
Hartle--Hawking vacuum \cite{Hartle:1976tp}, and its
decay can be studied in the Euclidean time formalism.\footnote{The
  Euclidean time approach does not work in non-equilibrium situations,
  such as e.g. a radiating BH in empty space described by the Unruh vacuum
  \cite{Unruh:1976db}. The method applicable in this general case has
  been developed in \cite{Shkerin:2021zbf,Shkerin:2021rhy}.}
For a spherically symmetric BH in $d$ dimensions 
this leads to the theory on a Euclidean
manifold ${\cal M}^2\times {\cal S}^{d-2}$, where ${\cal S}^{d-2}$ is
the $(d-2)$-dimensional sphere, and ${\cal M}^2$ has geometry of a
cigar 
with compactified Euclidean time $\tau$ playing the role of the
angular variable and the radial coordinate covering the region outside
the horizon \cite{Hartle:1976tp}.
This picture corresponds to the Euclidean partition function in the
Hartle--Hawking vacuum, and the tunneling solutions are saddles of
this partition function. 

Previous studies give rise to the following puzzle. Analysis of $O(3)$-symmetric configurations in 4d Schwarzschild background using the
thin-wall approximation yields only $\tau$-independent solutions of
the sphaleron type, with no analogs of periodic instantons
\cite{Arnold:1989cq}. Similarly, still
working within the thin-wall approximation but with the gravitational
back-reaction taken into account,
Refs.~\cite{Gregory:2013hja,Burda:2015yfa} found that the
dominant contribution to the vacuum decay probability around a static
BH is always provided by the sphaleron.
The absence of periodic instantons is surprising, 
and one may wonder if it is peculiar to the thin-wall regime. 

We provide evidence that this is not the case. Namely, we prove that
for a general multi-scalar theory and a wide class of spherically symmetric
BHs in $d$ dimensions the $\tau$-independent bounces centered on the BH have
exactly one $O(d-1)$-symmetric 
negative mode at any $T_{\rm BH}$. This makes
existence of any $O(d-1)$-symmetric periodic instantons around the BH
unlikely.\footnote{This, of course, does not rule out periodic
  instantons far away from the BH in the asymptotically flat region.} 
Throughout our analysis we neglect the back-reaction of the
scalar field on the metric. 

The paper is organized as follows. We first develop the intuition for
the proof on an example of 2-dimensional dilaton BH in
sec.~\ref{sec:2}. The proof is then extended to general
$d$-dimensional BHs in
sec.~\ref{sec:3}. We discuss our results in sec.~\ref{sec:4}.
We work in the system of units $\hbar=c=G=1$.

\section{Dilaton black hole in two dimensions}
\label{sec:2}

As discussed in \cite{Shkerin:2021zbf}, only the exterior region of a BH is relevant for the computation of the vacuum decay
probability. In this region the metric of a 2-dimensional dilaton BH
\cite{Callan:1992rs} can be written in the form,
\begin{equation}\label{ds}
\diff s^2=\O(x)(-\diff t^2+\diff x^2) \;,
\end{equation}
where $-\infty<x<\infty$ and 
\begin{equation}\label{O2}
\O=\frac{1}{1+\e^{-2\l x}} \;.
\end{equation}
The parameter $\l$ is related to the BH temperature,
$\l\equiv 2\pi T_{\rm BH}$. 
The horizon is located at $x\to-\infty$; in the opposite limit the
metric is asymptotically flat. 
We analytically continue the metric to the Euclidean time by
replacing $t\mapsto -i\tau$. The Euclidean time is compactified with
the period $2\pi/\l$ to ensure the absence of conical singularity at
$x=-\infty$.  

Consider a scalar field $\vf$ with the potential $V(\vf)$ embedded in
this metric.\footnote{The potential can depend on the
  temperature due to thermal loop corrections.}
The Euclidean action for $\vf$ reads,
\begin{equation}\label{S_2d}
S=\int\diff\tau\diff x\bigg\{ \frac{1}{2}\bigg(\frac{\d \vf}{\d x}\bigg)^2
+\frac{1}{2}\bigg(\frac{\d \vf}{\d
  \tau}\bigg)^2+\O V(\vf) \bigg\} \;.
\end{equation}
We assume that
$V(\vf)$ has a local metastable minimum $V=0$ at $\vf=0$. Then the
equation of motion for $\vf$ has a time-independent solution $\vf_s(x)$ ---
sphaleron --- which interpolates between $\vf=0$ at $x\to+\infty$ and
a nonzero value at the horizon. 
It describes the saddle point of the energy barrier
separating the false and true vacua. Note that the equation for $\vf$
contains the metric function $\Omega(x)$ and hence $\vf_s$
parametrically depends on the BH temperature $\l$. 

\begin{figure}[t]
	\center{	
		\begin{minipage}[h]{0.99\linewidth}
			\center{\includegraphics[width=0.9\linewidth]{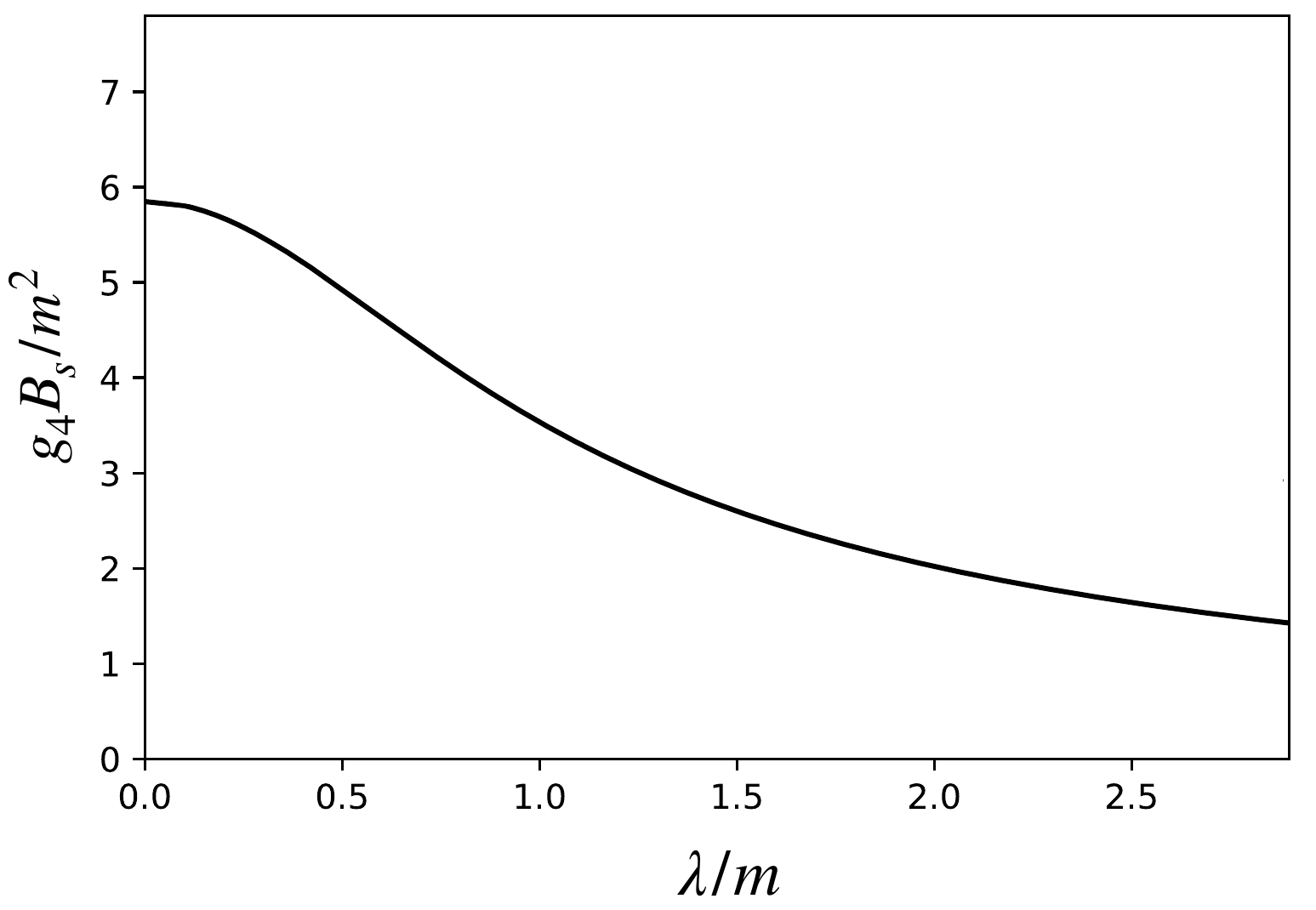}}
		\end{minipage}
	}
	\caption{Exponential suppression of sphalerons in the
          background of $2$-dimensional dilaton black hole in equilibrium with
          thermal bath for the theory with the potential
          (\ref{V_2d}).}
	\label{fig:2d_B_dil}
\end{figure}

The sphaleron gives rise to a $\tau$-independent bounce. The first
indication that this bounce dominates the vacuum decay at any
$\l$ comes from considering its Euclidean action $B_s$. 
For the model with the potential (\ref{V_2d}) it is
plotted in fig.~\ref{fig:2d_B_dil} which must be contrasted with
fig.~\ref{fig:2d_B_flat} in flat space.  
At high temperature the behaviour of the sphaleron suppression is
qualitatively similar in both cases. However, at low temperature it is
dramatically different. Instead of diverging at small $\l$ as its
flat-space counterpart, $B_s$ in the case of BH 
smoothly increases towards a finite
value  $B_0$ equal to the suppression of the flat-space vacuum
bounce. This is indeed the expected behaviour of the false vacuum
decay rate at low temperature.

To see what happens in more detail, let us focus on the near-horizon
region, $x<0$ and $\l|x|\gg 1$. The conformal factor is then approximated as
$\O\approx\e^{2\l x}$. 
We substitute this expression for $\O$ into (\ref{ds}) and make the
following change of variables 
\begin{equation}\label{Rindler}
	\Tau=\frac{1}{\l}\e^{\l x}\sin\l\tau \;, ~~~ X=\frac{1}{\l}\e^{\l x}\cos\l\tau \;.
\end{equation}
This brings the line element to the form $\diff s^2=\diff \Tau^2+\diff X^2$.
Thus, close to the BH horizon, the spacetime geometry is approximately flat.
The coordinates $(\tau,x)$ define the (Euclidean) Rindler frame in this flat space.
In the Lorentzian signature, the Rindler frame corresponds to
uniformly accelerating observers in the Minkowski spacetime. 
In particular, the line of constant $x$ represents a trajectory of an
observer moving with the acceleration $\l\e^{-\l x}$.  
The BH vicinity looks like the Rindler wedge---the
portion of the Minkowski spacetime accessible to the observer,
separated from the rest of the world by the horizon.

Consider the flat-space vacuum bounce $\vf_0(R)$, where
$R=\sqrt{\Tau^2+X^2}$, centered around the origin of
$(\Tau,X)$.\,\footnote{See \cite{0806.0299} for the proof of $O(2)$-symmetry of the vacuum bounce in 2 dimensions.} 
The transformation (\ref{Rindler}) converts it into a static solution
$\vf_0\left(\e^{\l x}/\l\right)$. 
Thus, the decay of the Minkowski vacuum in the Rindler frame is
mediated by a $\tau$-independent bounce or sphaleron. 
This agrees with the fact that the Minkowski vacuum corresponds to a
thermal state from the viewpoint of an accelerating observer
\cite{Ai:2018rnh,Shkerin:2021zbf}.
Note that though in Lorentzian signature 
the Rindler coordinates cover only half of the flat spatial slices, 
the Euclidean action of the bounce computed in these coordinates
coincides with the full flat-space integration in the Cartesian
coordinates and equals $B_0$. Returning to the BH sphaleron, 
its physical size is much smaller than
$\l^{-1}$ in the limit $\l\ll m$.
Hence it lies entirely in the near-horizon region and
coincides with the Rindler sphaleron, 
$\vf_s(x)=\vf_0\left(\e^{\l x}/\l\right)$. This explains the smooth
limit of its suppression $B_s\to B_0$ at $\l\to 0$.  

We can infer one more lesson from considering the Rindler wedge. In
Minkowski space centering bounce at the origin is not the only
possibility. We can as well translate it to an arbitrary point. In
particular, take the bounce centered at $\Tau=0$, $X=X_0$ and 
denote it by $\vf_{X_0}$. Switching to the Rindler frame converts it to a
$\tau$-dependent periodic instanton which nevertheless has exactly the
same Euclidean action $B_0$ as the sphaleron.    
In other words, the Rindler space possesses a degenerate family of
periodic Euclidean solutions---the `Rindler valley'---parameterized
by $X_0$, with the limit $X_0=0$ corresponding to the sphaleron.  
In the vicinity of the point $X_0=0$, the periodic instantons
from the valley are obtained from the sphaleron by a dipole perturbation
\begin{equation}\label{dipole_pert}
\vf_{X_0}\approx \vf_0 + X_0 \cos\theta\frac{\diff \vf_0}{\diff R} \;,
\end{equation}
where $\theta=\arctg\Tau/X$.
The deviation of the BH metric from Rindler tilts the valley. 
The direction of the tilt is then crucial. If it is towards 
$X_0=0$, the sphaleron is less suppressed than any of the
would-be periodic instantons. And vice versa, a tilt away from $X_0=0$
would lead to periodic instantons being less suppressed that the
sphaleron. We are going to see that the first option is realized,
ruling out periodic instantons connected to sphaleron. 

We presently prove
that the
$\tau$-independent bounce has only a single negative mode. 
Introducing polar coordinates
\begin{equation}\label{2dvar}
\s=\l^{-1}\arcsh(\e^{\l x}) \;, ~~~ \theta = \l\tau \;,
\end{equation}
the action
(\ref{S_2d}) takes the form
\begin{equation}\label{S_2d_new}
\begin{split}
S=\limitint_0^{2\pi}\diff\theta\limitint_0^\infty 
& \frac{\diff \s\th{\l \s}}{\l}
\bigg\{ \frac{1}{2}\bigg(\frac{\d\vf}{\d \s}\bigg)^2 \\
& +\frac{\l^2}{2\th^2\l \s}\bigg(\frac{\d\vf}{\d \theta}\bigg)^2 
+V(\vf)\bigg\}\;.
\end{split}
\end{equation} 
Note that the radial coordinate $\s$ is chosen to measure
the geodesic distance to the horizon which ensures that the term
$\tfrac{1}{2}(\d\vf/\d\s)^2$ appears in the action with the same
coefficient as the potential term.
The sphaleron has only radial dependence and satisfies
the equation,
\begin{equation}\label{ysph2d}
-\frac{\diff^2\vf_s}{\diff \s^2}-\frac{2\l}{\sh 2\l
  \s}\frac{\diff\vf_s}{\diff \s}+V'(\vf_s)=0\;,
\end{equation}
where prime means derivative of the potential with respect to the
field. 
Due to the rotational invariance, the eigenmodes of perturbations
around $\vf_s$ decompose into multipole sectors with the angular
dependence $\e^{\pm in\theta}$. A radial eigenfunction $\chi_\mu(\s)$ corresponding
to the eigenvalue $\mu$ in the $n$-th sector obeys a
Schr\"odinger-type equation
\begin{equation}\label{2dradial}
H_{{\rm eff},n}\chi_\mu=\mu\,\chi_\mu
\end{equation}
with the effective Hamiltonian 
\begin{equation}\label{Heff2d}
H_{{\rm eff},n}=
-\frac{\diff^2}{\diff \s^2}-\frac{2\l}{\sh 2\l \s}\frac{\diff}{\diff \s}
+ U_{{\rm eff},n}(\s)
\end{equation}
and 
\begin{equation}\label{Ueff2d}
U_{{\rm eff},n}(\s)=V''\big(\vf_s(\s)\big)+\frac{n^2\l^2}{\th^2\l \s}\;.
\end{equation}
Note that the Hamiltonian (\ref{Heff2d}) is Hermitian with respect to
the positive measure 
$\int \diff \s\, \l^{-1}\th\l \s$ and the potential (\ref{Ueff2d}) is
positive at $\s\to\infty$. Hence, negative modes (${\mu<0}$) of $H_{{\rm
    eff},n}$, if any, form a discrete spectrum. We will take them
to be normalized. 

First, note that in the monopole sector ($n=0$) the sphaleron has
exactly one negative mode. Indeed, monopole perturbations correspond
to static configurations whose Euclidean action is given simply by
their energy times $2\pi/\l$. But the sphaleron is, by definition, the
saddle-point of the static energy functional and hence the minimum of
the energy in all but one direction.

Next we show that the spectrum in the dipole sector ($n=1$) is
strictly positive. This is the key part of the proof. We take the
derivative of the sphaleron equation (\ref{ysph2d}) with respect to
$\sigma$ and obtain,
\begin{equation}\label{ydots2d}
-\frac{\diff^2\dot\vf_s}{\diff \s^2}-\frac{2\l}{\sh 2\l
  \s}\frac{\diff\dot\vf_s}{\diff \s}+\tilde U_{\rm eff}(\s)\dot\vf_s=0\;,
\end{equation}
where $\dot{\vf}_s\equiv \diff \vf_s/\diff \s$ 
and we have introduced a new effective potential
\begin{equation}\label{Ueff12d}
\tilde U_{{\rm eff}}(\s)= U_{{\rm eff},1}(\s)-\lambda^2\th^2 \l \s\;.
\end{equation}
This implies that $\dot\vf_s$ is a zero mode of the new Hamiltonian
\begin{equation}\label{Heff12d}
\tilde H_{{\rm eff}}=-\frac{\diff^2}{\diff \s^2}
-\frac{2\l}{\sh 2\l \s}\frac{\diff}{\diff \s}
+\tilde U_{{\rm eff}}(\s)\;.
\end{equation}
The sphaleron is a monotonic function of $\s$, as can be
easily shown using the analogy between \cref{ysph2d} and the equation
of motion of a unit-mass particle moving with the friction in the
potential $-V(\vf_s)$ \cite{Coleman:1977py}. 
This means that $\dot\vf_s$ does not cross zero and, hence, it is the
ground state of $\tilde H_{\rm eff}$.  
But $H_{{\rm eff},1}$ differs from $\tilde H_{{\rm eff}}$ by an
addition of a positive potential $\l^2\th^2\l \s$, and therefore, its
discrete eigenvalues are strictly larger than those of $\tilde H_{{\rm
    eff}}$.\,\footnote{To prove this, consider a continuous
  deformation of $\tilde H_{{\rm eff}}$ into $H_{{\rm eff},1}$ by
  monotonically increasing the potential and use
  $\delta\mu=\bra{\chi_\mu}\delta H\ket{\chi_\mu}$.}   
Thus, the ground state energy of $H_{{\rm eff},1}$ is strictly
positive. 

Clearly, the last result also implies the positivity of the spectrum
in higher multipoles since $U_{{\rm eff},n}(\s)>U_{{\rm eff},1}(\s)$
for $n>1$. Thus, we conclude that the only negative mode of the
$\tau$-independent bounce resides in the monopole sector. 
This completes the proof.

The above proof is
straightforwardly generalized to the theory of several scalar fields
$\vf^{i}$.  
Proceeding as before, one obtains two Hamiltonians: $H_{{\rm eff},1}$
for the perturbations in the dipole sector, and $\tilde H_{{\rm eff}}$
arising upon differentiation of the sphaleron equation. 
The difference between $H_{{\rm eff},1}$ and
$\tilde H_{{\rm eff}}$ is diagonal in the space of fields and is manifestly
positive.  
Hence, the eigenvalues of $H_{{\rm eff},1}$ are strictly larger than
those of $\tilde H_{{\rm eff}}$.  
On the other hand, the latter has the zero mode $\dot \vf_s^i$. 
The only thing that needs to be proven is that $\dot \vf_s^i$ is the
ground state of $\tilde H_{{\rm eff}}$.  
Consider the low-temperature limit in which, as we know, the sphaleron
is obtained from
the flat vacuum bounce $\vf_0^i(\s)$ by the Rindler
transformation. 
Since $\vf_0^i$ is a tunneling solution, it has a single negative mode
in the monopole sector, and no negative modes in the dipole sector.
Also $\tilde H_{{\rm eff}}=H_{{\rm eff},1}$ at $\l=0$.
Hence, $\dot \vf_s^i$ is the ground state of $\tilde H_{{\rm eff}}$ 
at small temperature and,
by continuity, it remains to be the ground state at all temperatures.

\section{Black holes in $d$ dimensions}
\label{sec:3}

We are ready to address the
case of a spherically symmetric BH in $d>2$ dimensions. 
Consider the metric
\begin{equation}\label{4dmetric}
\diff s^2=f(r)\diff\tau^2+\frac{\diff r^2}{f(r)}+r^2\diff\Sigma_{d-2}^2\;,
\end{equation}
where $\diff\Sigma_{d-2}^2$ is the line element on the
$(d-2)$-dimensional unit sphere. 
We do not specify $f(r)$ at this point, assuming only that it has
horizon at $r=r_h$, where $f(r_h)=0$, $f'(r_h)>0$, and that it is
positive at $r>r_h$. 
We analyze explicitly the theory of a single real scalar field; the
multi-field generalization is straightforward, as in the 2d
case studied above. 

A theory with a metastable vacuum has a sphaleron solution that inherits the  
$O(d-1)$ symmetry of the metric. We want to prove that, for a broad
class of BHs,
it has a single
negative mode within the sector of $O(d-1)$-symmetric
configurations. 
For such configurations the Euclidean action reads,  
\begin{equation}\label{S_4d}
\begin{split}
S&={\cal A}_{d-2}\int \diff\tau \diff r\,r^{d-2}
\bigg\{\frac{f}{2}\bigg(\frac{\d\vf}{\d r}\bigg)^2
+\frac{1}{2f}\bigg(\frac{\d\vf}{\d \tau}\bigg)^2+V(\vf)\bigg\}\\
&=\frac{2{\cal A}_{d-2}}{f'(r_h)}
\int \diff \theta \diff \s\,r^{d-2} 
\sqrt{f}\bigg\{\frac{1}{2}\bigg(\frac{\d\vf}{\d \s}\bigg)^2 \\
&\qquad\qquad+\frac{(f'(r_h))^2}{8f}\bigg(\frac{\d\vf}{\d \theta}\bigg)^2+V(\vf)\bigg\} \;,
\end{split}
\end{equation}
where ${\cal A}_{d-2}=2\pi^{\frac{d-1}{2}}/\Gamma(\frac{d-1}{2})$ is
the surface area of a unit $(d-2)$-dimensional sphere, and in the
second line we introduced the geodesic coordinates 
\begin{equation}\label{4dytau}
\s=\int\frac{\diff r}{\sqrt{f(r)}}~,~~~~\theta=\frac{f'(r_h)}{2}\tau\;.
\end{equation}
Note that $\s$ is set to be zero at the horizon, $\s(r_h)=0$, and
$\theta$ is an angular variable with period $2\pi$ ensuring the
regularity of the Euclidean metric (\ref{4dmetric}) at the horizon. 

As in 2 dimensions, it is sufficient to focus on the dipole perturbations of
the form $\chi_\mu(\s)\cos\theta$. The radial function satisfies the
equation 
\begin{equation}\label{chi04d}
-\frac{\diff^2\chi_\mu}{\diff
  \s^2}-\frac{\diff\ln(r^{d-2}\sqrt{f})}{\diff
  \s}\frac{\diff\chi_\mu}{\diff \s}+ 
U_{{\rm eff},1}(\s)\chi_\mu =\mu\,\chi_\mu\;
\end{equation}
with the effective potential
\begin{equation}\label{Ueff4d}
U_{{\rm eff},1}=V''(\vf_s)+\frac{(f'(r_h))^2}{4f}\;.
\end{equation}
On the other hand, by taking the $\s$-derivative of the sphaleron
equation 
\begin{equation}\label{4dsph}
-\frac{\diff^2\vf_s}{\diff \s^2}-\frac{\diff\ln(r^{d-2}\sqrt{f})}{\diff \s}\frac{\diff\vf_s}{\diff \s}+V'(\vf_s)=0\;,
\end{equation} 
we obtain
\begin{equation}\label{4ddotsph}
-\frac{\diff ^2\dot\vf_s}{\diff \s^2}
-\frac{\diff\ln(r^{d-2}\sqrt{f})}{\diff \s}\frac{\diff\dot\vf_s}{\diff
  \s}
+\tilde U_{{\rm eff}}(\s)\dot\vf_s=0\;.
\end{equation}
As before, we have defined $\dot{\vf}_s\equiv\diff\vf_s/\diff \s$ and
\begin{equation}\label{Ueff14d}
\begin{split}
& \tilde U_{{\rm eff}} =V''(\vf_s)-\frac{\diff^2\ln(r^{d-2}\sqrt{f})}{\diff \s^2} \\
&=V''(\vf_s)+\frac{(d-2)f}{r^2}-\frac{(d-2)f'}{2r}-\frac{f''}{2}+\frac{(f')^2}{4f}\;,
\end{split}
\end{equation}
where $f'$ stands for the derivative of $f$ with respect to $r$. 
We observe that $\dot\vf_s$ is the ground state\footnote{$\dot\vf_s$
  does not have nodes as long as $\vf_s(\s)$ is monotonic. A
  sufficient condition for this is the positivity of the ``friction''
  term in \cref{4dsph}, $\diff\ln(r^{d-2}\sqrt{f})/\diff\s>0$.} 
with zero energy of
the Hamiltonian $\tilde H_{{\rm eff}}$ corresponding to the potential
(\ref{Ueff14d}). Hence, \cref{chi04d} does not have any negative
eigenmodes if 
$U_{{\rm eff},1}>\tilde U_{{\rm eff}}$, or equivalently if
\begin{equation}\label{4dineq}
\begin{split}
D(r)\equiv
&\frac{(f'(r_h))^2-(f')^2}{4}\\
&-\frac{(d-2)f^2}{r^2}+\frac{(d-2)ff'}{2r}+\frac{ff''}{2}>0\;
\end{split}
\end{equation}
at $r>r_h$. This inequality provides a general sufficient condition
for the absence of non-monopole negative modes on the $\tau$-independent
bounce. Note that it involves only the properties of the metric, but
not of the scalar potential.

We now show that the condition (\ref{4dineq}) is satisfied for several
common BH metrics. We start with  
the Schwarzschild metric, $f(r)=1-2M/r^{d-3}$.
Instead of computing $D(r)$ directly, it is easier to find its
derivative,
\be
\label{DSch}
D'(r)=\frac{2(d-2)}{r^3}\left(1-\frac{(d-1)M}{r^{d-3}}\right)^2\;.
\ee
We see that the latter is positive. Since $D(r_h)=0$, this implies
positivity of $D(r)$ everywhere outside the horizon. The argument extends to the 
Schwarzschild-(anti) de Sitter metric,\footnote{The Schwarzschild-de
  Sitter metric has, apart from the BH horizon, a cosmological
  horizon. We consider here a thermal ensemble in equilibrium with the
former, but not the latter.} $f(r)=1-2M/r^{d-3}\pm r^2/l^2$.
Remarkably, $l^2$ drops out of $D'(r)$ which is
still given by (\ref{DSch}).

Next consider the Reissner--Nordstr\"om BH with
\begin{equation}\label{RS}
f(r)=1-\frac{2M}{r^{d-3}}+\frac{Q^2}{r^{2(d-3)}} \;,
\end{equation}
where $M$ and $Q$ are BH mass and charge,
respectively. 
Inequality (\ref{4dineq}) is satisfied outside the horizon as long as
$|Q|$ is small enough. 
For example, in 4 dimensions a numerical analysis yields the 
bound $Q^2<Q_*^2$, where $Q_*^2\simeq 0.833\,M^2$. 

On the other hand, a nearly critical BH provides a notable violation
of the condition (\ref{4dineq}). Computing $D(r)$ for $Q=M$ we obtain,
\be
\label{DRS}
D(r)=-\frac{d-2}{r^2}\left(1-\frac{M}{r^{d-3}}\right)^4<0\;.
\ee 
Recalling that $D(r)$ sets the difference between $U_{{\rm
    eff},1}$ and $\tilde U_{\rm eff}$, we conclude that in this case the
$\tau$-independent bounce {\em does} have a negative dipole
mode. Hence, for a nearly critical BH, it is not a valid solution for
the false vacuum decay and we expect existence of periodic instantons
centered on the BH. 

We focused above on $O(d-1)$-symmetric configurations. Our results do
not exclude existence of extra negative modes which break this
symmetry. In fact, presence of such negative modes is expected for
large enough BHs. Indeed, consider the equation for a mode which is
$\tau$-independent, but has non-trivial dependence on the angles
$\Sigma_{d-2}$,
\be
-\frac{\diff^2\chi_\mu}{\diff\s^2}-\frac{\diff\ln(r^{d-2}\sqrt{f})}{\diff\s}
\frac{\diff\chi_\mu}{\diff\s}
+\left(\!\!V''(\vf_s)+\frac{L^2}{r^2}\right)\chi_\mu=\mu\,\chi_\mu 
\,.
\ee 
Here $-L^2$ is the eigenvalue of the Laplacian on a unit
$(d-2)$-dimensional sphere. We know that for $L^2=0$ this equation has
a negative eigenvalue $\mu_-<0$. Then, since $r$ is bounded from below
by $r_h$, the equations with non-zero $L^2$ also have negative modes
if $L^2/r_h^2<|\mu_-|$. This instability has an intuitive
explanation. 
For large
$r_h$ the symmetric sphaleron looks like a shell stretched on the BH horizon
and locally has almost flat geometry. Not surprisingly,
it is unstable with respect to inhomogeneous perturbations. Note that
in this regime the sphaleron suppression is proportional to the
horizon area $\propto r_h^{d-2}$ and exceeds the suppression of flat-space
bounces. Since large BHs have low temperature $T_{\rm BH}\lesssim
r_h^{-1}$, we conclude that low-temperature decays are dominated by
bounces away from the BH. This is an important difference of the $d>2$
case from the theory in $2$ dimensions.

\section{Discussion}
\label{sec:4}

We have shown for a large class of BHs that 
the $\tau$-independent $O(d-1)$-symmetric bounce centered on them
has a single negative mode in the sector
of $O(d-1)$-symmetric perturbations. This negative mode is
$\tau$-independent. Though this does not strictly rule out a branch of
$O(d-1)$-symmetric periodic instantons completely disconnected from
the sphaleron, we believe this possibility to be
unlikely. Asymmetric, but still $\tau$-independent, negative modes do
exist for large (and cold) BHs. In this case, however, the bounce
centered on the BH is more suppressed than the bounce in the
asymptotically flat region and is irrelevant for vacuum decay.

Our result has two implications. On the technical side, it simplifies
the analysis of BH catalysis of false vacuum decay in thermal bath 
by sparing the task
of looking for periodic instantons. The latter requires solving
nonlinear elliptic partial differential equations, or a system thereof
for multi-field theories. By contrast, the static sphaleron depends
only on the radial coordinate and is found as a solution of ordinary
differential equations. This simplification can be useful in various
contexts where one deals with thermal ensembles containing BHs, such
as e.g. cosmology \cite{Gorbunov:2017fhq,Canko:2017ebb,Mukaida:2017bgd} or holography
\cite{Creminelli:2001th,Freivogel:2005qh,Agashe:2019lhy,Bigazzi:2020phm}.  
 
From the physical perspective, the dominance of $\tau$-independent
bounce suggests to interpret false vacuum decay around the BH as
driven by over-barrier transitions at any $T_{\rm BH}$. 
This appears consistent
with the fact that, from the viewpoint of the asymptotic observer, the
temperature in the vicinity of the BH is blue-shifted and exceeds
$T_c$ close enough to the horizon leading to enhanced thermal
fluctuations. However, the clear distinction between quantum tunneling
and over-barrier transitions is absent in curved spacetime. This is
known from the study of Coleman--De Luccia \cite{Coleman:1980aw} and 
Hawking--Moss instantons \cite{Hawking:1981fz} in de Sitter
which can be interpreted either as vacuum bounces
in inflationary coordinates
\cite{Rubakov:1999ir}, or as thermal sphaleron transitions in the static
patch \cite{Brown:2007sd}. Another example is provided by the relation between
Minkowski and Rindler bounces 
\cite{Ai:2018rnh,Shkerin:2021zbf,Ai:2022kqm} mentioned in
sec.~\ref{sec:2}. Similarly, the static bounce in
Schwarzschild metric becomes time-dependent in a different reference
frame, such as e.g. Kruskal coordinates, and in this frame corresponds
to a
vacuum bounce. Thus, the BH spacetime gives one more example that the
question whether the vacuum decay is due to quantum or thermal
fluctuations has no covariant meaning.

A notable exception to our proof is provided by nearly critical
Reissner--Nordstr\"om BH, in which case we expect existence of
periodic instantons centered on the BH. An intuitive explanation why
such BHs are special comes from considering an exactly critical
BH. This has zero temperature and the state around it appears as
vacuum even for a static observer. The true bounce in this case must
be localized in the Euclidean time, as in Minkowski space, whereas the
$\tau$-independent bounce has infinite suppression.  

In this paper we focused on thermal ensemble and did not consider
non-equilibrium situations, such as a BH emitting Hawking radiation into
empty space (Unruh vacuum). It will be interesting to see if the
approach of this work can be used to get insight into the
structure of complex bounces describing false vacuum decay in this
case \cite{Shkerin:2021zbf,Shkerin:2021rhy}. Finally, it would be
interesting to extend our analysis by including gravitational
back-reaction of the bounce on the metric.

\noindent{\bf\emph{Acknowledgments.}}---We thank Ruth Gregory and Matthew Johnson for
useful discussions. The work of A.S. is supported in part by the
Department of Energy under Grant No. DE-SC0011842. 
The work of S.S. is supported by
the Natural Sciences and Engineering Research Council (NSERC) of Canada.
Research at Perimeter Institute is supported in part by the Government
of Canada through the Department of Innovation, Science and Economic
Development Canada and by the Province of Ontario through the Ministry
of Colleges and Universities.

\bibliography{Refs}

\begin{thebibliography}{40}%
\makeatletter
\providecommand \@ifxundefined [1]{%
 \@ifx{#1\undefined}
}%
\providecommand \@ifnum [1]{%
 \ifnum #1\expandafter \@firstoftwo
 \else \expandafter \@secondoftwo
 \fi
}%
\providecommand \@ifx [1]{%
 \ifx #1\expandafter \@firstoftwo
 \else \expandafter \@secondoftwo
 \fi
}%
\providecommand \natexlab [1]{#1}%
\providecommand \enquote  [1]{``#1''}%
\providecommand \bibnamefont  [1]{#1}%
\providecommand \bibfnamefont [1]{#1}%
\providecommand \citenamefont [1]{#1}%
\providecommand \href@noop [0]{\@secondoftwo}%
\providecommand \href [0]{\begingroup \@sanitize@url \@href}%
\providecommand \@href[1]{\@@startlink{#1}\@@href}%
\providecommand \@@href[1]{\endgroup#1\@@endlink}%
\providecommand \@sanitize@url [0]{\catcode `\\12\catcode `\$12\catcode
  `\&12\catcode `\#12\catcode `\^12\catcode `\_12\catcode `\%12\relax}%
\providecommand \@@startlink[1]{}%
\providecommand \@@endlink[0]{}%
\providecommand \url  [0]{\begingroup\@sanitize@url \@url }%
\providecommand \@url [1]{\endgroup\@href {#1}{\urlprefix }}%
\providecommand \urlprefix  [0]{URL }%
\providecommand \Eprint [0]{\href }%
\providecommand \doibase [0]{http://dx.doi.org/}%
\providecommand \selectlanguage [0]{\@gobble}%
\providecommand \bibinfo  [0]{\@secondoftwo}%
\providecommand \bibfield  [0]{\@secondoftwo}%
\providecommand \translation [1]{[#1]}%
\providecommand \BibitemOpen [0]{}%
\providecommand \bibitemStop [0]{}%
\providecommand \bibitemNoStop [0]{.\EOS\space}%
\providecommand \EOS [0]{\spacefactor3000\relax}%
\providecommand \BibitemShut  [1]{\csname bibitem#1\endcsname}%
\let\auto@bib@innerbib\@empty
\bibitem [{\citenamefont {Weinberg}(2012)}]{Weinberg:2012pjx}%
  \BibitemOpen
  \bibfield  {author} {\bibinfo {author} {\bibfnamefont {E.~J.}\ \bibnamefont
  {Weinberg}},\ }\href {\doibase 10.1017/CBO9781139017787} {\emph {\bibinfo
  {title} {{Classical solutions in quantum field theory}: {Solitons and
  Instantons in High Energy Physics}}}},\ Cambridge Monographs on Mathematical
  Physics\ (\bibinfo  {publisher} {Cambridge University Press},\ \bibinfo
  {year} {2012})\BibitemShut {NoStop}%
\bibitem [{\citenamefont {Coleman}(1977)}]{Coleman:1977py}%
  \BibitemOpen
  \bibfield  {author} {\bibinfo {author} {\bibfnamefont {S.~R.}\ \bibnamefont
  {Coleman}},\ }\href {\doibase 10.1103/PhysRevD.15.2929,
  10.1103/PhysRevD.16.1248} {\bibfield  {journal} {\bibinfo  {journal} {Phys.
  Rev.}\ }\textbf {\bibinfo {volume} {D15}},\ \bibinfo {pages} {2929} (\bibinfo
  {year} {1977})},\ \bibinfo {note} {[Erratum: Phys.
  Rev.D16,1248(1977)]}\BibitemShut {NoStop}%
\bibitem [{\citenamefont {Callan}\ and\ \citenamefont
  {Coleman}(1977)}]{Callan:1977pt}%
  \BibitemOpen
  \bibfield  {author} {\bibinfo {author} {\bibfnamefont {C.~G.}\ \bibnamefont
  {Callan}, \bibfnamefont {Jr.}}\ and\ \bibinfo {author} {\bibfnamefont
  {S.~R.}\ \bibnamefont {Coleman}},\ }\href {\doibase 10.1103/PhysRevD.16.1762}
  {\bibfield  {journal} {\bibinfo  {journal} {Phys. Rev.}\ }\textbf {\bibinfo
  {volume} {D16}},\ \bibinfo {pages} {1762} (\bibinfo {year}
  {1977})}\BibitemShut {NoStop}%
\bibitem [{\citenamefont {Linde}(1981)}]{Linde:1980tt}%
  \BibitemOpen
  \bibfield  {author} {\bibinfo {author} {\bibfnamefont {A.~D.}\ \bibnamefont
  {Linde}},\ }\href {\doibase 10.1016/0370-2693(81)90281-1} {\bibfield
  {journal} {\bibinfo  {journal} {Phys. Lett.}\ }\textbf {\bibinfo {volume}
  {100B}},\ \bibinfo {pages} {37} (\bibinfo {year} {1981})}\BibitemShut
  {NoStop}%
\bibitem [{\citenamefont {Linde}(1983)}]{Linde:1981zj}%
  \BibitemOpen
  \bibfield  {author} {\bibinfo {author} {\bibfnamefont {A.~D.}\ \bibnamefont
  {Linde}},\ }\href {\doibase 10.1016/0550-3213(83)90072-X} {\bibfield
  {journal} {\bibinfo  {journal} {Nucl. Phys. B}\ }\textbf {\bibinfo {volume}
  {216}},\ \bibinfo {pages} {421} (\bibinfo {year} {1983})},\ \bibinfo {note}
  {[Erratum: Nucl.Phys.B 223, 544 (1983)]}\BibitemShut {NoStop}%
\bibitem [{\citenamefont {Garriga}(1994)}]{Garriga:1994ut}%
  \BibitemOpen
  \bibfield  {author} {\bibinfo {author} {\bibfnamefont {J.}~\bibnamefont
  {Garriga}},\ }\href {\doibase 10.1103/PhysRevD.49.5497} {\bibfield  {journal}
  {\bibinfo  {journal} {Phys. Rev. D}\ }\textbf {\bibinfo {volume} {49}},\
  \bibinfo {pages} {5497} (\bibinfo {year} {1994})},\ \Eprint
  {http://arxiv.org/abs/hep-th/9401020} {arXiv:hep-th/9401020} \BibitemShut
  {NoStop}%
\bibitem [{\citenamefont {Hiscock}(1987)}]{Hiscock:1987hn}%
  \BibitemOpen
  \bibfield  {author} {\bibinfo {author} {\bibfnamefont {W.~A.}\ \bibnamefont
  {Hiscock}},\ }\href {\doibase 10.1103/PhysRevD.35.1161} {\bibfield  {journal}
  {\bibinfo  {journal} {Phys. Rev.}\ }\textbf {\bibinfo {volume} {D35}},\
  \bibinfo {pages} {1161} (\bibinfo {year} {1987})}\BibitemShut {NoStop}%
\bibitem [{\citenamefont {Berezin}\ \emph {et~al.}(1988)\citenamefont
  {Berezin}, \citenamefont {Kuzmin},\ and\ \citenamefont
  {Tkachev}}]{Berezin:1987ea}%
  \BibitemOpen
  \bibfield  {author} {\bibinfo {author} {\bibfnamefont {V.~A.}\ \bibnamefont
  {Berezin}}, \bibinfo {author} {\bibfnamefont {V.~A.}\ \bibnamefont {Kuzmin}},
  \ and\ \bibinfo {author} {\bibfnamefont {I.~I.}\ \bibnamefont {Tkachev}},\
  }\href {\doibase 10.1016/0370-2693(88)90672-7} {\bibfield  {journal}
  {\bibinfo  {journal} {Phys. Lett.}\ }\textbf {\bibinfo {volume} {B207}},\
  \bibinfo {pages} {397} (\bibinfo {year} {1988})}\BibitemShut {NoStop}%
\bibitem [{\citenamefont {Arnold}(1990)}]{Arnold:1989cq}%
  \BibitemOpen
  \bibfield  {author} {\bibinfo {author} {\bibfnamefont {P.~B.}\ \bibnamefont
  {Arnold}},\ }\href {\doibase 10.1016/0550-3213(90)90243-7} {\bibfield
  {journal} {\bibinfo  {journal} {Nucl. Phys.}\ }\textbf {\bibinfo {volume}
  {B346}},\ \bibinfo {pages} {160} (\bibinfo {year} {1990})}\BibitemShut
  {NoStop}%
\bibitem [{\citenamefont {Berezin}\ \emph {et~al.}(1991)\citenamefont
  {Berezin}, \citenamefont {Kuzmin},\ and\ \citenamefont
  {Tkachev}}]{Berezin:1990qs}%
  \BibitemOpen
  \bibfield  {author} {\bibinfo {author} {\bibfnamefont {V.~A.}\ \bibnamefont
  {Berezin}}, \bibinfo {author} {\bibfnamefont {V.~A.}\ \bibnamefont {Kuzmin}},
  \ and\ \bibinfo {author} {\bibfnamefont {I.~I.}\ \bibnamefont {Tkachev}},\
  }\href {\doibase 10.1103/PhysRevD.43.R3112} {\bibfield  {journal} {\bibinfo
  {journal} {Phys. Rev.}\ }\textbf {\bibinfo {volume} {D43}},\ \bibinfo {pages}
  {3112} (\bibinfo {year} {1991})}\BibitemShut {NoStop}%
\bibitem [{\citenamefont {Gregory}\ \emph {et~al.}(2014)\citenamefont
  {Gregory}, \citenamefont {Moss},\ and\ \citenamefont
  {Withers}}]{Gregory:2013hja}%
  \BibitemOpen
  \bibfield  {author} {\bibinfo {author} {\bibfnamefont {R.}~\bibnamefont
  {Gregory}}, \bibinfo {author} {\bibfnamefont {I.~G.}\ \bibnamefont {Moss}}, \
  and\ \bibinfo {author} {\bibfnamefont {B.}~\bibnamefont {Withers}},\ }\href
  {\doibase 10.1007/JHEP03(2014)081} {\bibfield  {journal} {\bibinfo  {journal}
  {JHEP}\ }\textbf {\bibinfo {volume} {03}},\ \bibinfo {pages} {081} (\bibinfo
  {year} {2014})},\ \Eprint {http://arxiv.org/abs/1401.0017} {arXiv:1401.0017
  [hep-th]} \BibitemShut {NoStop}%
\bibitem [{\citenamefont {Sasaki}\ and\ \citenamefont
  {Yeom}(2014)}]{Sasaki:2014spa}%
  \BibitemOpen
  \bibfield  {author} {\bibinfo {author} {\bibfnamefont {M.}~\bibnamefont
  {Sasaki}}\ and\ \bibinfo {author} {\bibfnamefont {D.-h.}\ \bibnamefont
  {Yeom}},\ }\href {\doibase 10.1007/JHEP12(2014)155} {\bibfield  {journal}
  {\bibinfo  {journal} {JHEP}\ }\textbf {\bibinfo {volume} {12}},\ \bibinfo
  {pages} {155} (\bibinfo {year} {2014})},\ \Eprint
  {http://arxiv.org/abs/1404.1565} {arXiv:1404.1565 [hep-th]} \BibitemShut
  {NoStop}%
\bibitem [{\citenamefont {Burda}\ \emph
  {et~al.}(2015{\natexlab{a}})\citenamefont {Burda}, \citenamefont {Gregory},\
  and\ \citenamefont {Moss}}]{Burda:2015isa}%
  \BibitemOpen
  \bibfield  {author} {\bibinfo {author} {\bibfnamefont {P.}~\bibnamefont
  {Burda}}, \bibinfo {author} {\bibfnamefont {R.}~\bibnamefont {Gregory}}, \
  and\ \bibinfo {author} {\bibfnamefont {I.}~\bibnamefont {Moss}},\ }\href
  {\doibase 10.1103/PhysRevLett.115.071303} {\bibfield  {journal} {\bibinfo
  {journal} {Phys. Rev. Lett.}\ }\textbf {\bibinfo {volume} {115}},\ \bibinfo
  {pages} {071303} (\bibinfo {year} {2015}{\natexlab{a}})},\ \Eprint
  {http://arxiv.org/abs/1501.04937} {arXiv:1501.04937 [hep-th]} \BibitemShut
  {NoStop}%
\bibitem [{\citenamefont {Burda}\ \emph
  {et~al.}(2015{\natexlab{b}})\citenamefont {Burda}, \citenamefont {Gregory},\
  and\ \citenamefont {Moss}}]{Burda:2015yfa}%
  \BibitemOpen
  \bibfield  {author} {\bibinfo {author} {\bibfnamefont {P.}~\bibnamefont
  {Burda}}, \bibinfo {author} {\bibfnamefont {R.}~\bibnamefont {Gregory}}, \
  and\ \bibinfo {author} {\bibfnamefont {I.}~\bibnamefont {Moss}},\ }\href
  {\doibase 10.1007/JHEP08(2015)114} {\bibfield  {journal} {\bibinfo  {journal}
  {JHEP}\ }\textbf {\bibinfo {volume} {08}},\ \bibinfo {pages} {114} (\bibinfo
  {year} {2015}{\natexlab{b}})},\ \Eprint {http://arxiv.org/abs/1503.07331}
  {arXiv:1503.07331 [hep-th]} \BibitemShut {NoStop}%
\bibitem [{\citenamefont {Burda}\ \emph {et~al.}(2016)\citenamefont {Burda},
  \citenamefont {Gregory},\ and\ \citenamefont {Moss}}]{Burda:2016mou}%
  \BibitemOpen
  \bibfield  {author} {\bibinfo {author} {\bibfnamefont {P.}~\bibnamefont
  {Burda}}, \bibinfo {author} {\bibfnamefont {R.}~\bibnamefont {Gregory}}, \
  and\ \bibinfo {author} {\bibfnamefont {I.}~\bibnamefont {Moss}},\ }\href
  {\doibase 10.1007/JHEP06(2016)025} {\bibfield  {journal} {\bibinfo  {journal}
  {JHEP}\ }\textbf {\bibinfo {volume} {06}},\ \bibinfo {pages} {025} (\bibinfo
  {year} {2016})},\ \Eprint {http://arxiv.org/abs/1601.02152} {arXiv:1601.02152
  [hep-th]} \BibitemShut {NoStop}%
\bibitem [{\citenamefont {Tetradis}(2016)}]{Tetradis:2016vqb}%
  \BibitemOpen
  \bibfield  {author} {\bibinfo {author} {\bibfnamefont {N.}~\bibnamefont
  {Tetradis}},\ }\href {\doibase 10.1088/1475-7516/2016/09/036} {\bibfield
  {journal} {\bibinfo  {journal} {JCAP}\ }\textbf {\bibinfo {volume} {1609}},\
  \bibinfo {pages} {036} (\bibinfo {year} {2016})},\ \Eprint
  {http://arxiv.org/abs/1606.04018} {arXiv:1606.04018 [hep-ph]} \BibitemShut
  {NoStop}%
\bibitem [{\citenamefont {Chen}\ \emph {et~al.}(2017)\citenamefont {Chen},
  \citenamefont {Dom\`enech}, \citenamefont {Sasaki},\ and\ \citenamefont
  {Yeom}}]{Chen:2017suz}%
  \BibitemOpen
  \bibfield  {author} {\bibinfo {author} {\bibfnamefont {P.}~\bibnamefont
  {Chen}}, \bibinfo {author} {\bibfnamefont {G.}~\bibnamefont {Dom\`enech}},
  \bibinfo {author} {\bibfnamefont {M.}~\bibnamefont {Sasaki}}, \ and\ \bibinfo
  {author} {\bibfnamefont {D.-h.}\ \bibnamefont {Yeom}},\ }\href {\doibase
  10.1007/JHEP07(2017)134} {\bibfield  {journal} {\bibinfo  {journal} {JHEP}\
  }\textbf {\bibinfo {volume} {07}},\ \bibinfo {pages} {134} (\bibinfo {year}
  {2017})},\ \Eprint {http://arxiv.org/abs/1704.04020} {arXiv:1704.04020
  [gr-qc]} \BibitemShut {NoStop}%
\bibitem [{\citenamefont {Gorbunov}\ \emph {et~al.}(2017)\citenamefont
  {Gorbunov}, \citenamefont {Levkov},\ and\ \citenamefont
  {Panin}}]{Gorbunov:2017fhq}%
  \BibitemOpen
  \bibfield  {author} {\bibinfo {author} {\bibfnamefont {D.}~\bibnamefont
  {Gorbunov}}, \bibinfo {author} {\bibfnamefont {D.}~\bibnamefont {Levkov}}, \
  and\ \bibinfo {author} {\bibfnamefont {A.}~\bibnamefont {Panin}},\ }\href
  {\doibase 10.1088/1475-7516/2017/10/016} {\bibfield  {journal} {\bibinfo
  {journal} {JCAP}\ }\textbf {\bibinfo {volume} {1710}},\ \bibinfo {pages}
  {016} (\bibinfo {year} {2017})},\ \Eprint {http://arxiv.org/abs/1704.05399}
  {arXiv:1704.05399 [astro-ph.CO]} \BibitemShut {NoStop}%
\bibitem [{\citenamefont {Canko}\ \emph {et~al.}(2018)\citenamefont {Canko},
  \citenamefont {Gialamas}, \citenamefont {Jelic-Cizmek}, \citenamefont
  {Riotto},\ and\ \citenamefont {Tetradis}}]{Canko:2017ebb}%
  \BibitemOpen
  \bibfield  {author} {\bibinfo {author} {\bibfnamefont {D.}~\bibnamefont
  {Canko}}, \bibinfo {author} {\bibfnamefont {I.}~\bibnamefont {Gialamas}},
  \bibinfo {author} {\bibfnamefont {G.}~\bibnamefont {Jelic-Cizmek}}, \bibinfo
  {author} {\bibfnamefont {A.}~\bibnamefont {Riotto}}, \ and\ \bibinfo {author}
  {\bibfnamefont {N.}~\bibnamefont {Tetradis}},\ }\href {\doibase
  10.1140/epjc/s10052-018-5808-y} {\bibfield  {journal} {\bibinfo  {journal}
  {Eur. Phys. J.}\ }\textbf {\bibinfo {volume} {C78}},\ \bibinfo {pages} {328}
  (\bibinfo {year} {2018})},\ \Eprint {http://arxiv.org/abs/1706.01364}
  {arXiv:1706.01364 [hep-th]} \BibitemShut {NoStop}%
\bibitem [{\citenamefont {Mukaida}\ and\ \citenamefont
  {Yamada}(2017)}]{Mukaida:2017bgd}%
  \BibitemOpen
  \bibfield  {author} {\bibinfo {author} {\bibfnamefont {K.}~\bibnamefont
  {Mukaida}}\ and\ \bibinfo {author} {\bibfnamefont {M.}~\bibnamefont
  {Yamada}},\ }\href {\doibase 10.1103/PhysRevD.96.103514} {\bibfield
  {journal} {\bibinfo  {journal} {Phys. Rev.}\ }\textbf {\bibinfo {volume}
  {D96}},\ \bibinfo {pages} {103514} (\bibinfo {year} {2017})},\ \Eprint
  {http://arxiv.org/abs/1706.04523} {arXiv:1706.04523 [hep-th]} \BibitemShut
  {NoStop}%
\bibitem [{\citenamefont {Kohri}\ and\ \citenamefont
  {Matsui}(2018)}]{Kohri:2017ybt}%
  \BibitemOpen
  \bibfield  {author} {\bibinfo {author} {\bibfnamefont {K.}~\bibnamefont
  {Kohri}}\ and\ \bibinfo {author} {\bibfnamefont {H.}~\bibnamefont {Matsui}},\
  }\href {\doibase 10.1103/PhysRevD.98.123509} {\bibfield  {journal} {\bibinfo
  {journal} {Phys. Rev.}\ }\textbf {\bibinfo {volume} {D98}},\ \bibinfo {pages}
  {123509} (\bibinfo {year} {2018})},\ \Eprint
  {http://arxiv.org/abs/1708.02138} {arXiv:1708.02138 [hep-ph]} \BibitemShut
  {NoStop}%
\bibitem [{\citenamefont {Hayashi}\ \emph {et~al.}(2020)\citenamefont
  {Hayashi}, \citenamefont {Kamada}, \citenamefont {Oshita},\ and\
  \citenamefont {Yokoyama}}]{Hayashi:2020ocn}%
  \BibitemOpen
  \bibfield  {author} {\bibinfo {author} {\bibfnamefont {T.}~\bibnamefont
  {Hayashi}}, \bibinfo {author} {\bibfnamefont {K.}~\bibnamefont {Kamada}},
  \bibinfo {author} {\bibfnamefont {N.}~\bibnamefont {Oshita}}, \ and\ \bibinfo
  {author} {\bibfnamefont {J.}~\bibnamefont {Yokoyama}},\ }\href {\doibase
  10.1007/JHEP08(2020)088} {\bibfield  {journal} {\bibinfo  {journal} {JHEP}\
  }\textbf {\bibinfo {volume} {08}},\ \bibinfo {pages} {088} (\bibinfo {year}
  {2020})},\ \Eprint {http://arxiv.org/abs/2005.12808} {arXiv:2005.12808
  [hep-th]} \BibitemShut {NoStop}%
\bibitem [{\citenamefont {Miyachi}\ and\ \citenamefont
  {Soda}(2021)}]{Miyachi:2021bwd}%
  \BibitemOpen
  \bibfield  {author} {\bibinfo {author} {\bibfnamefont {T.}~\bibnamefont
  {Miyachi}}\ and\ \bibinfo {author} {\bibfnamefont {J.}~\bibnamefont {Soda}},\
  }\href {\doibase 10.1103/PhysRevD.103.085009} {\bibfield  {journal} {\bibinfo
   {journal} {Phys. Rev. D}\ }\textbf {\bibinfo {volume} {103}},\ \bibinfo
  {pages} {085009} (\bibinfo {year} {2021})},\ \Eprint
  {http://arxiv.org/abs/2102.02462} {arXiv:2102.02462 [gr-qc]} \BibitemShut
  {NoStop}%
\bibitem [{\citenamefont {Shkerin}\ and\ \citenamefont
  {Sibiryakov}(2021)}]{Shkerin:2021zbf}%
  \BibitemOpen
  \bibfield  {author} {\bibinfo {author} {\bibfnamefont {A.}~\bibnamefont
  {Shkerin}}\ and\ \bibinfo {author} {\bibfnamefont {S.}~\bibnamefont
  {Sibiryakov}},\ }\href {\doibase 10.1007/JHEP11(2021)197} {\bibfield
  {journal} {\bibinfo  {journal} {JHEP}\ }\textbf {\bibinfo {volume} {11}},\
  \bibinfo {pages} {197} (\bibinfo {year} {2021})},\ \Eprint
  {http://arxiv.org/abs/2105.09331} {arXiv:2105.09331 [hep-th]} \BibitemShut
  {NoStop}%
\bibitem [{\citenamefont {Shkerin}\ and\ \citenamefont
  {Sibiryakov}(2022)}]{Shkerin:2021rhy}%
  \BibitemOpen
  \bibfield  {author} {\bibinfo {author} {\bibfnamefont {A.}~\bibnamefont
  {Shkerin}}\ and\ \bibinfo {author} {\bibfnamefont {S.}~\bibnamefont
  {Sibiryakov}},\ }\href {\doibase 10.1007/JHEP08(2022)161} {\bibfield
  {journal} {\bibinfo  {journal} {JHEP}\ }\textbf {\bibinfo {volume} {08}},\
  \bibinfo {pages} {161} (\bibinfo {year} {2022})},\ \Eprint
  {http://arxiv.org/abs/2111.08017} {arXiv:2111.08017 [hep-th]} \BibitemShut
  {NoStop}%
\bibitem [{\citenamefont {Strumia}(2022)}]{Strumia:2022jil}%
  \BibitemOpen
  \bibfield  {author} {\bibinfo {author} {\bibfnamefont {A.}~\bibnamefont
  {Strumia}},\ }\href@noop {} {\enquote {\bibinfo {title} {{Black holes don't
  source fast Higgs vacuum decay}},}\ } (\bibinfo {year} {2022}),\ \Eprint
  {http://arxiv.org/abs/2209.05504} {arXiv:2209.05504 [hep-ph]} \BibitemShut
  {NoStop}%
\bibitem [{\citenamefont {Hartle}\ and\ \citenamefont
  {Hawking}(1976)}]{Hartle:1976tp}%
  \BibitemOpen
  \bibfield  {author} {\bibinfo {author} {\bibfnamefont {J.~B.}\ \bibnamefont
  {Hartle}}\ and\ \bibinfo {author} {\bibfnamefont {S.~W.}\ \bibnamefont
  {Hawking}},\ }\href {\doibase 10.1103/PhysRevD.13.2188} {\bibfield  {journal}
  {\bibinfo  {journal} {Phys. Rev.}\ }\textbf {\bibinfo {volume} {D13}},\
  \bibinfo {pages} {2188} (\bibinfo {year} {1976})}\BibitemShut {NoStop}%
\bibitem [{\citenamefont {Unruh}(1976)}]{Unruh:1976db}%
  \BibitemOpen
  \bibfield  {author} {\bibinfo {author} {\bibfnamefont {W.~G.}\ \bibnamefont
  {Unruh}},\ }\href {\doibase 10.1103/PhysRevD.14.870} {\bibfield  {journal}
  {\bibinfo  {journal} {Phys. Rev.}\ }\textbf {\bibinfo {volume} {D14}},\
  \bibinfo {pages} {870} (\bibinfo {year} {1976})}\BibitemShut {NoStop}%
\bibitem [{\citenamefont {Callan}\ \emph {et~al.}(1992)\citenamefont {Callan},
  \citenamefont {Giddings}, \citenamefont {Harvey},\ and\ \citenamefont
  {Strominger}}]{Callan:1992rs}%
  \BibitemOpen
  \bibfield  {author} {\bibinfo {author} {\bibfnamefont {C.~G.}\ \bibnamefont
  {Callan}, \bibfnamefont {Jr.}}, \bibinfo {author} {\bibfnamefont {S.~B.}\
  \bibnamefont {Giddings}}, \bibinfo {author} {\bibfnamefont {J.~A.}\
  \bibnamefont {Harvey}}, \ and\ \bibinfo {author} {\bibfnamefont
  {A.}~\bibnamefont {Strominger}},\ }\href {\doibase 10.1103/PhysRevD.45.R1005}
  {\bibfield  {journal} {\bibinfo  {journal} {Phys. Rev.}\ }\textbf {\bibinfo
  {volume} {D45}},\ \bibinfo {pages} {R1005} (\bibinfo {year} {1992})},\
  \Eprint {http://arxiv.org/abs/hep-th/9111056} {arXiv:hep-th/9111056 [hep-th]}
  \BibitemShut {NoStop}%
\bibitem [{\citenamefont {{Byeon}}\ \emph {et~al.}(2008)\citenamefont
  {{Byeon}}, \citenamefont {{Jeanjean}},\ and\ \citenamefont
  {{Mari{\c{s}}}}}]{0806.0299}%
  \BibitemOpen
  \bibfield  {author} {\bibinfo {author} {\bibfnamefont {J.}~\bibnamefont
  {{Byeon}}}, \bibinfo {author} {\bibfnamefont {L.}~\bibnamefont {{Jeanjean}}},
  \ and\ \bibinfo {author} {\bibfnamefont {M.}~\bibnamefont {{Mari{\c{s}}}}},\
  }\href@noop {} {\enquote {\bibinfo {title} {{Symmetry and monotonicity of
  least energy solutions}},}\ } (\bibinfo {year} {2008}),\ \Eprint
  {http://arxiv.org/abs/0806.0299} {arXiv:0806.0299 [math.AP]} \BibitemShut
  {NoStop}%
\bibitem [{\citenamefont {Ai}(2019)}]{Ai:2018rnh}%
  \BibitemOpen
  \bibfield  {author} {\bibinfo {author} {\bibfnamefont {W.-Y.}\ \bibnamefont
  {Ai}},\ }\href {\doibase 10.1007/JHEP03(2019)164} {\bibfield  {journal}
  {\bibinfo  {journal} {JHEP}\ }\textbf {\bibinfo {volume} {03}},\ \bibinfo
  {pages} {164} (\bibinfo {year} {2019})},\ \Eprint
  {http://arxiv.org/abs/1812.06962} {arXiv:1812.06962 [hep-th]} \BibitemShut
  {NoStop}%
\bibitem [{\citenamefont {Creminelli}\ \emph {et~al.}(2002)\citenamefont
  {Creminelli}, \citenamefont {Nicolis},\ and\ \citenamefont
  {Rattazzi}}]{Creminelli:2001th}%
  \BibitemOpen
  \bibfield  {author} {\bibinfo {author} {\bibfnamefont {P.}~\bibnamefont
  {Creminelli}}, \bibinfo {author} {\bibfnamefont {A.}~\bibnamefont {Nicolis}},
  \ and\ \bibinfo {author} {\bibfnamefont {R.}~\bibnamefont {Rattazzi}},\
  }\href {\doibase 10.1088/1126-6708/2002/03/051} {\bibfield  {journal}
  {\bibinfo  {journal} {JHEP}\ }\textbf {\bibinfo {volume} {03}},\ \bibinfo
  {pages} {051} (\bibinfo {year} {2002})},\ \Eprint
  {http://arxiv.org/abs/hep-th/0107141} {arXiv:hep-th/0107141} \BibitemShut
  {NoStop}%
\bibitem [{\citenamefont {Freivogel}\ \emph {et~al.}(2006)\citenamefont
  {Freivogel}, \citenamefont {Hubeny}, \citenamefont {Maloney}, \citenamefont
  {Myers}, \citenamefont {Rangamani},\ and\ \citenamefont
  {Shenker}}]{Freivogel:2005qh}%
  \BibitemOpen
  \bibfield  {author} {\bibinfo {author} {\bibfnamefont {B.}~\bibnamefont
  {Freivogel}}, \bibinfo {author} {\bibfnamefont {V.~E.}\ \bibnamefont
  {Hubeny}}, \bibinfo {author} {\bibfnamefont {A.}~\bibnamefont {Maloney}},
  \bibinfo {author} {\bibfnamefont {R.~C.}\ \bibnamefont {Myers}}, \bibinfo
  {author} {\bibfnamefont {M.}~\bibnamefont {Rangamani}}, \ and\ \bibinfo
  {author} {\bibfnamefont {S.}~\bibnamefont {Shenker}},\ }\href {\doibase
  10.1088/1126-6708/2006/03/007} {\bibfield  {journal} {\bibinfo  {journal}
  {JHEP}\ }\textbf {\bibinfo {volume} {03}},\ \bibinfo {pages} {007} (\bibinfo
  {year} {2006})},\ \Eprint {http://arxiv.org/abs/hep-th/0510046}
  {arXiv:hep-th/0510046} \BibitemShut {NoStop}%
\bibitem [{\citenamefont {Agashe}\ \emph {et~al.}(2020)\citenamefont {Agashe},
  \citenamefont {Du}, \citenamefont {Ekhterachian}, \citenamefont {Kumar},\
  and\ \citenamefont {Sundrum}}]{Agashe:2019lhy}%
  \BibitemOpen
  \bibfield  {author} {\bibinfo {author} {\bibfnamefont {K.}~\bibnamefont
  {Agashe}}, \bibinfo {author} {\bibfnamefont {P.}~\bibnamefont {Du}}, \bibinfo
  {author} {\bibfnamefont {M.}~\bibnamefont {Ekhterachian}}, \bibinfo {author}
  {\bibfnamefont {S.}~\bibnamefont {Kumar}}, \ and\ \bibinfo {author}
  {\bibfnamefont {R.}~\bibnamefont {Sundrum}},\ }\href {\doibase
  10.1007/JHEP05(2020)086} {\bibfield  {journal} {\bibinfo  {journal} {JHEP}\
  }\textbf {\bibinfo {volume} {05}},\ \bibinfo {pages} {086} (\bibinfo {year}
  {2020})},\ \Eprint {http://arxiv.org/abs/1910.06238} {arXiv:1910.06238
  [hep-ph]} \BibitemShut {NoStop}%
\bibitem [{\citenamefont {Bigazzi}\ \emph {et~al.}(2020)\citenamefont
  {Bigazzi}, \citenamefont {Caddeo}, \citenamefont {Cotrone},\ and\
  \citenamefont {Paredes}}]{Bigazzi:2020phm}%
  \BibitemOpen
  \bibfield  {author} {\bibinfo {author} {\bibfnamefont {F.}~\bibnamefont
  {Bigazzi}}, \bibinfo {author} {\bibfnamefont {A.}~\bibnamefont {Caddeo}},
  \bibinfo {author} {\bibfnamefont {A.~L.}\ \bibnamefont {Cotrone}}, \ and\
  \bibinfo {author} {\bibfnamefont {A.}~\bibnamefont {Paredes}},\ }\href
  {\doibase 10.1007/JHEP12(2020)200} {\bibfield  {journal} {\bibinfo  {journal}
  {JHEP}\ }\textbf {\bibinfo {volume} {12}},\ \bibinfo {pages} {200} (\bibinfo
  {year} {2020})},\ \Eprint {http://arxiv.org/abs/2008.02579} {arXiv:2008.02579
  [hep-th]} \BibitemShut {NoStop}%
\bibitem [{\citenamefont {Coleman}\ and\ \citenamefont
  {De~Luccia}(1980)}]{Coleman:1980aw}%
  \BibitemOpen
  \bibfield  {author} {\bibinfo {author} {\bibfnamefont {S.~R.}\ \bibnamefont
  {Coleman}}\ and\ \bibinfo {author} {\bibfnamefont {F.}~\bibnamefont
  {De~Luccia}},\ }\href {\doibase 10.1103/PhysRevD.21.3305} {\bibfield
  {journal} {\bibinfo  {journal} {Phys. Rev.}\ }\textbf {\bibinfo {volume}
  {D21}},\ \bibinfo {pages} {3305} (\bibinfo {year} {1980})}\BibitemShut
  {NoStop}%
\bibitem [{\citenamefont {Hawking}\ and\ \citenamefont
  {Moss}(1982)}]{Hawking:1981fz}%
  \BibitemOpen
  \bibfield  {author} {\bibinfo {author} {\bibfnamefont {S.~W.}\ \bibnamefont
  {Hawking}}\ and\ \bibinfo {author} {\bibfnamefont {I.~G.}\ \bibnamefont
  {Moss}},\ }\href {\doibase 10.1016/0370-2693(82)90946-7} {\bibfield
  {journal} {\bibinfo  {journal} {Phys. Lett. B}\ }\textbf {\bibinfo {volume}
  {110}},\ \bibinfo {pages} {35} (\bibinfo {year} {1982})}\BibitemShut
  {NoStop}%
\bibitem [{\citenamefont {Rubakov}\ and\ \citenamefont
  {Sibiryakov}(1999)}]{Rubakov:1999ir}%
  \BibitemOpen
  \bibfield  {author} {\bibinfo {author} {\bibfnamefont {V.~A.}\ \bibnamefont
  {Rubakov}}\ and\ \bibinfo {author} {\bibfnamefont {S.~M.}\ \bibnamefont
  {Sibiryakov}},\ }\href {\doibase 10.1007/BF02557243} {\bibfield  {journal}
  {\bibinfo  {journal} {Theor. Math. Phys.}\ }\textbf {\bibinfo {volume}
  {120}},\ \bibinfo {pages} {1194} (\bibinfo {year} {1999})},\ \bibinfo {note}
  {[Teor. Mat. Fiz.120,451(1999)]},\ \Eprint
  {http://arxiv.org/abs/gr-qc/9905093} {arXiv:gr-qc/9905093 [gr-qc]}
  \BibitemShut {NoStop}%
\bibitem [{\citenamefont {Brown}\ and\ \citenamefont
  {Weinberg}(2007)}]{Brown:2007sd}%
  \BibitemOpen
  \bibfield  {author} {\bibinfo {author} {\bibfnamefont {A.~R.}\ \bibnamefont
  {Brown}}\ and\ \bibinfo {author} {\bibfnamefont {E.~J.}\ \bibnamefont
  {Weinberg}},\ }\href {\doibase 10.1103/PhysRevD.76.064003} {\bibfield
  {journal} {\bibinfo  {journal} {Phys. Rev. D}\ }\textbf {\bibinfo {volume}
  {76}},\ \bibinfo {pages} {064003} (\bibinfo {year} {2007})},\ \Eprint
  {http://arxiv.org/abs/0706.1573} {arXiv:0706.1573 [hep-th]} \BibitemShut
  {NoStop}%
\bibitem [{\citenamefont {Ai}\ \emph {et~al.}(2022)\citenamefont {Ai},
  \citenamefont {Cruz}, \citenamefont {Garbrecht},\ and\ \citenamefont
  {Tamarit}}]{Ai:2022kqm}%
  \BibitemOpen
  \bibfield  {author} {\bibinfo {author} {\bibfnamefont {W.-Y.}\ \bibnamefont
  {Ai}}, \bibinfo {author} {\bibfnamefont {J.~S.}\ \bibnamefont {Cruz}},
  \bibinfo {author} {\bibfnamefont {B.}~\bibnamefont {Garbrecht}}, \ and\
  \bibinfo {author} {\bibfnamefont {C.}~\bibnamefont {Tamarit}},\ }\href@noop
  {} {\enquote {\bibinfo {title} {{Instability of bubble expansion at zero
  temperature}},}\ } (\bibinfo {year} {2022}),\ \Eprint
  {http://arxiv.org/abs/2209.00639} {arXiv:2209.00639 [hep-th]} \BibitemShut
  {NoStop}%
\end{thebibliography}%

\end{document}